\newcommand{\be}{\begin{equation}}
\newcommand{\ee}{\end{equation}}

\documentclass[aps,prl,twocolumn,superscriptaddress]{revtex4}
\usepackage{graphicx}
\usepackage{amssymb}
\usepackage{bm}

\begin{document}
\title{Band-gap solitons in nonlinear optically-induced lattices}

\author{Anton S. Desyatnikov}
\affiliation{Nonlinear Photonics Group, Institute of Applied
Physics, Westf\"{a}lische Wilhelms-Universit\"{a}t M\"{u}nster, D-48149
M\"{u}nster, Germany} \affiliation{Nonlinear Physics Group, Research
School of Physical Sciences and Engineering, Australian National
University, Canberra ACT 0200, Australia}
\author{Elena A. Ostrovskaya}
\affiliation{Nonlinear Physics Group, Research School of Physical
Sciences and Engineering, Australian National University, Canberra
ACT 0200, Australia}
\author{Yuri S. Kivshar}
\affiliation{Nonlinear Physics Group, Research School of Physical
Sciences and Engineering, Australian National University, Canberra
ACT 0200, Australia}
\author{Cornelia Denz}
\affiliation{Nonlinear Photonics Group, Institute of Applied
Physics, Westf\"{a}lische Wilhelms-Universit\"{a}t M\"{u}nster, D-48149
M\"{u}nster, Germany}

\begin{abstract}
We introduce novel optical solitons that consist of a periodic and
a spatially localized components coupled nonlinearly via
cross-phase modulation. The spatially localized optical field can
be treated as a gap soliton supported by the optically-induced
nonlinear grating. We find different types of these band-gap
composite solitons and demonstrate their dynamical stability.
\end{abstract}

\maketitle

Recent theoretical and experimental results demonstrated nonlinear
localization of light in optically-induced refractive index
gratings \cite{efrem,fleischer}. Such localized states can be
treated as ``discrete'' and ``gap'' solitons observed in
fabricated periodic photonic structures \cite{book}, but supported
by gratings induced by a complementary optical field. Optically
induced lattices open up an exciting possibility for creating
dynamically reconfigurable photonic structures in bulk nonlinear
media. The physics of coherent light propagating in such
structures can be linked to the phenomena exhibited by coherent
matter waves (Bose-Einstein condensates) in optical lattices
\cite{lenaPRA}.

Among the most challenging problems in the physics of induced
gratings is the creation of {\em stable, uniform periodic optical
patterns} which can effectively modulate the refractive index of a
nonlinear medium. Periodic modulation of the refractive index can
be induced, for instance, by an interference pattern illuminating
a photorefractive crystal with a strong electro-optic anisotropy
\cite{efrem}. Interfering plane waves modulate the space-charge
field in the crystal, which relates to the refractive index via
electro-optic coefficients. The latter are substantially different
for the two orthogonal polarizations. As a result, the {\em
material nonlinearity} experienced by waves polarized in the
direction of the {\em c} - axis of the crystal is up to two orders
of magnitude larger than that experienced by the orthogonally
polarized ones. When the lattice-forming waves are polarized
orthogonally to the {\em c} - axis, the nonlinear self-action as
well as any cross-action from the co-propagating probe beam can be
neglected. The periodic interference pattern propagates in the
diffraction-free {\em linear} regime, thus creating a stationary
refractive-index grating \cite{fleischer}.

In this Letter, we develop the concept of optically-induced
gratings beyond the limit of weak material nonlinearity and
propose the idea of robust {\em nonlinearity-assisted optical
lattices}, created by {\em nonlinear periodic waves}. Strong
incoherent interaction of such a grating with a probe beam,
through the nonlinear cross-phase-modulation (XPM) effect,
facilitates the formation of a novel type of a composite optical
soliton, where one of the components creates a periodic photonic
structure, while the other component experiences Bragg reflection
from this structure and can form gap solitons localized in the
transmission gaps of the linear spectrum. The observation of
nonlinear light localization in this type of optically-induced
gratings can be achieved in photorefractive medium with two
incoherently interacting beams of the {\em same polarization}. We
study such a configuration in a saturable medium and demonstrate
the existence and stable dynamics of these novel band-gap lattice
solitons.

The propagation of two incoherently interacting beams in a
photorefractive crystal can be approximately described by the
coupled nonlinear Schr\"odinger (NLS) equations for the slowly
varying envelopes $E_n$ ($n=1,2$),
\be \label{NLS} i \frac{\partial E_n}{\partial z} + \frac{
\partial^2 E_n}{\partial x^2} + \sigma N(I) E_n=0, \ee
where $N(I)= I/(1+sI)$ describes saturable nonlinearity, $I=\sum
|E_n|^2$ is the total light intensity, $s$ is the saturation
parameter, and $\sigma=\pm 1$ stands for the focusing or
defocusing nonlinearity, respectively. Stationary solutions are
found in the form $E_n=u_n(x) \exp(i\sigma k_nz)$ where $k_n$ are
the propagation constants of the components. We assume strong
saturation regime, $s=1$, which is closer to realistic
experimental conditions.

\begin{figure}
\centerline{\scalebox{0.75}{\includegraphics{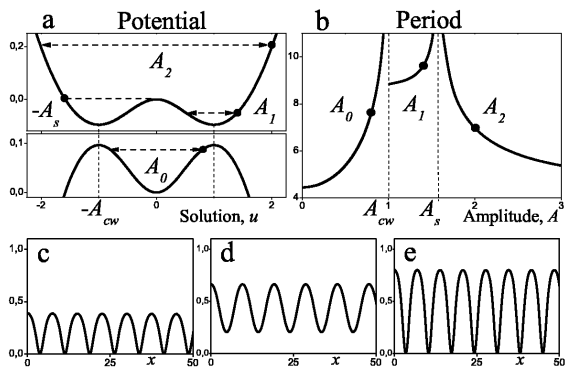}}}\vspace{-1mm}
\caption{(a) Effective potential $P(u_1)$ for the focusing (top)
and defocusing (bottom) cases ($k_1=0.5$). Period (b) and examples
(c-e) of the refractive index modulation $N(I)$ for the three
branches of periodic solutions of the NLS equation: (c) $A_0=0.8$,
(d) $A_1=1.41$, and (e) $A_2=2.01$.} \label{figpot}
\end{figure}

The {\em induced waveguiding} regime, well studied in the context
of vector solitons \cite{book}, corresponds to the case when the
intensities of the two interacting fields are significantly
different. Then the strong field (e.g., $u_1$) is described by a
single (scalar) NLS equation, and the weaker field propagates in
the effective {\em linear waveguide} induced by the stronger
component via XPM. Here we assume that the effective waveguide
(i.e. the grating) is created by a periodic nonlinear field
$u_1(x)$  with the propagation constant $k_1$, described by the
stationary wave-train solutions of a scalar equation (\ref{NLS})
(see also Ref.~\cite{kart}). Integrating the stationary form of
Eq.~(\ref{NLS}) once, we introduce the effective potential
$P(u_1)= (\sigma/2) [(1-k_1)u^2_1-\ln(1+u^2_1)]$, so that the
general stationary solution $u_1(x)$ with the amplitude $A$ can be
found by solving the equation
$P(A)=\frac{1}{2}(du_1/dx)^2+P(u_1)$. Figure \ref{figpot}(a) shows
the form of the potential $P(u_1)$ in both focusing ($\sigma =+1$,
top) and defocusing ($\sigma =-1$, bottom) cases. The minima of
$P(u_1)$ correspond to a plane wave with the constant amplitude
$A^2_{\rm cw}=k_1/(1-k_1)$, whereas the bright soliton solutions
correspond to the separatrix at $A=A_s$ with $P(A_s)=0$.

\begin{figure}
\centerline{\scalebox{0.8}{\includegraphics{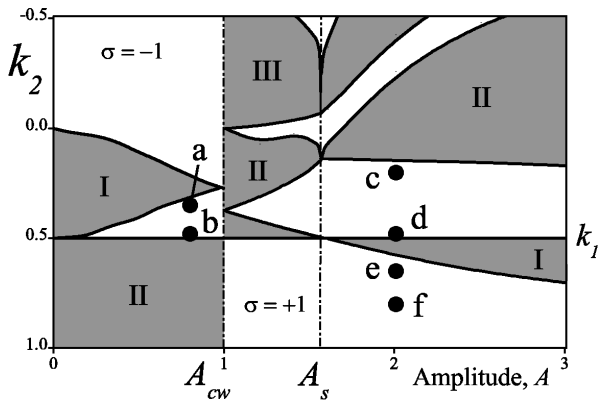}}}\vspace{-1mm}
\caption{The bandgap structure of the linear spectrum $-k_2(A)$,
induced by the nonlinear periodic grating $u_1(x)$ for $k_1=0.5$.
The bands are shaded. Marked dots indicate the propagation
constants for gap solitons shown in Fig.~\ref{figsol}.}
\label{figBG}
\end{figure}

In the limit $s\to 0$, exact analytical expressions for the
nonlinear periodic waves $u_1(x)$ can be written down for
$\sigma=\pm1$, in terms of the elliptic Jacobi functions, ${\rm
cn}(x,\mu)$, ${\rm dn}(x,\mu)$, and ${\rm sn}(x,\mu)$, with the
modulus $0\leq \mu(k_1)\leq 1$. They represent self-consistent
solutions of the cubic NLS equations which coincide with
well-studied Hill's equation with associated Lam{\'e} potentials
\cite{petn}. It can be shown that the general structure of the
periodic solutions is preserved for $s\neq 0$. For $\sigma =+1$,
there exist two branches of the periodic (cnoidal) solutions shown
through their induced refractive index modulation in
Fig.~\ref{figpot}(d), for $A=A_1$, $A_{\rm cw}<A_1<A_s$, and in
Fig.~\ref{figpot}(e), for $A=A_2>A_s$. The $cn$-type solutions of
the branch $A_2$ have nodes, whereas the $dn$-type solutions of
the branch $A_1$ are nodeless. In the defocusing case
($\sigma=-1$), there exists only one branch of the $sn$-type
periodic solutions for $A=A_0 < A_{\rm cw}$, see
Fig.~\ref{figpot}(c). In a strongly nonlinear limit the
large-period $A_{1,2}$ and $A_0$ solutions describe periodic
trains of bright and dark solitons, respectively.

Having identified the stationary, nonlinear periodic solutions for
the scalar field $u_1(x)$, we find that, in the induced
waveguiding regime, the weak wave $u_2$ is scattered by an {\em
effectively fixed linear grating} characterized by the potential
$N(I)$, where $I=u^2_1(x)$. The guiding properties of such a {\em
linear grating} are determined by the bandgap structure of the
spectrum of the Hill's equation: $d^2 u_2/dx^2=-\sigma
N(I)u_2+k_2u_2$, where the eigenvalue $k_2(A)$ depends on the
grating amplitude. The eigenfunctions satisfy the Bloch condition
$u_2(x)=\exp(iKL)u_2(x+L)$, where $L$ is the period and $K$ is the
momentum of the lattice. The spectrum consists of $M$ bands and a
continuum band, with the total $m=2M+1$ band edges. The
eigenfunction at the $m$-th band edge corresponds to a strictly
periodic Bloch wave $u^m_2 \equiv b^m(x)=\pm b^m(x+L)$, for which
$KL=0,\pi$. Figure \ref{figBG} shows an example of the bandgap
spectrum $-k_2$ generated by the scalar cnoidal wave $u_1(x)$, for
$k_1=0.5$. The Bloch waves at the band edges $m=1,2,3,4,5,\dots$
have the propagation constants $-\sigma k^m_2\leq -\sigma
k^{m+1}_2$ and the period $L,2L,2L,L,L,\dots$. The Bloch wave
$b^m(x)$ at the band edge $k^m_2=k_1$ coincides with the scalar
cnoidal wave $u_1(x)$. In the case of a saturable nonlinearity
some predictions of the number and position of the bands and gaps
can be made using theory of Lam{\'e}-type equations \cite{lame_b}.
For example, in the case of a defocusing nonlinearity, the grating
potential can be well approximated by $N(x)\approx a(a+1)\mu {\rm
sn}^2(x,\mu)$, where $a=1$. Since the spectrum of the Lam{\'e}'s
equation has $M=a$ bound bands, it is expected that the $A_0$-type
grating generates single bound band followed by a semi-infinite
band (as seen in Fig.~\ref{figBG} for $A<A_{\rm cw}$).

In the limit $A\rightarrow A_s$ the period of the cnoidal-wave
solution diverges and the spectrum bands disappear, see
Fig.~\ref{figBG}. From the other hand, when the grating amplitude
$A$ approaches the plane-wave amplitude $A_{\rm cw}$ in the
focusing case, the periodic modulation of the refractive index
vanishes and the gaps disappear, see Fig.~\ref{figBG}.

To be useful for creation of robust dynamical photonic structures,
nonlinear periodic waves should be stable. Previous studies of
stability of periodic solutions \cite{kart,stab,bec1d} suggest
that the solutions of the $A_1$-type are strongly unstable due to
modulational instability (MI), whereas MI is suppressed for the
$A_2$-type solutions in a saturable medium, and also for
$A_0$-type solutions in the defocusing case. Our numerical studies
have confirmed that the $A_0$-type grating in the defocusing case
is both linearly and dynamically stable, and also demonstrated
that the $A_2$-type solutions are only weakly (oscillatory)
unstable. In contrast, the $A_1$-type lattice is quickly destroyed
by strong symmetry-breaking instabilities; therefore we excluded
it from our further consideration.

\begin{figure}
\centerline{\scalebox{0.8}{\includegraphics{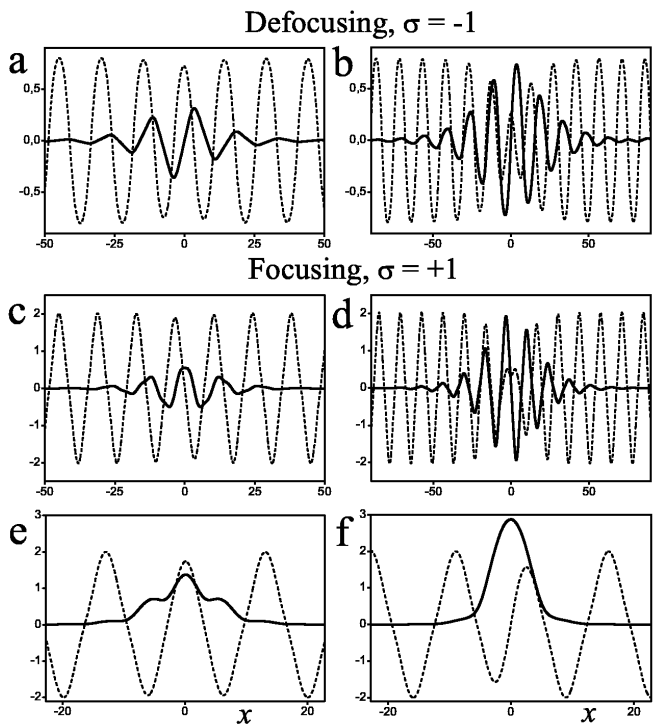}}}\vspace{-1mm}
\caption{Examples of the stationary two-component solutions in the
different bandgaps for fixed parameters of the periodical
component $k_1=0.5$ for the defocusing medium (the upper row) and
the focusing medium (two lower rows). Dashed - the periodic
component $u_1$; solid - the localized component $u_2$. Gap
solitons correspond to the marked points in Fig.~\ref{figpow}.}
\label{figsol}
\end{figure}

The localization of the probe field $u_2$ in the gaps of the
linear spectrum of the periodic structure induced by the field
$u_1(x)$ can occur in the {\em nonlinear} regime of the probe
propagation through the grating. In this regime, the significant
intensity of the probe beam does not permit to neglect its
nonlinear self-action. When the back-action of the probe on the
grating through XPM is ignored (e.g. in the case of a weak
material nonlinearity for the grating wave), the physics of the
localization is similar to the standard case of nonlinear waves in
fixed periodic potentials, well studied in the context of both
optical and matter waves \cite{sipe,andrey,bec1d,lenaPRA}.
However, in our problem the grating and scattered wave are {\em
strongly nonlinearly coupled} and, therefore, as in the case of
two-component vector solitons \cite{book}, we should expect the
existence of self-consistent hybrid structures formed by a
periodic wave and a localized gap mode.

\begin{figure}
\centerline{\scalebox{0.8}{\includegraphics{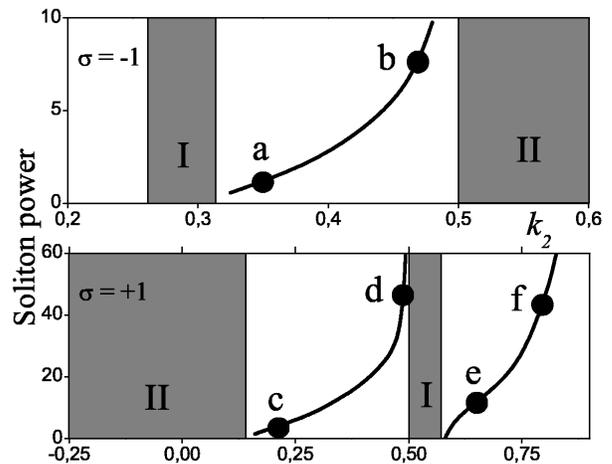}}}
\caption{Power of the localized component, $Q=\int u^2_2(x) dx$,
vs. $k_2$ for $k_1=0.5$ and (top) defocusing ($A_0=0.8$) and
(bottom) focusing ($A_2=2.01$) nonlinearity. Shaded regions
corresponds to the spectral bands, and the dots mark the solutions
shown in Fig.~\ref{figsol}.} \label{figpow}
\end{figure}

Indeed, by solving vector Eq. (\ref{NLS}) numerically, with a
value of $k_1$ {\em fixed} to that of the scalar grating
$k_1=k^g_1$, we have found different families of solutions of Eq.
(\ref{NLS}), consisting of the oscillatory ($u_1$) and localized
($u_2$) mutually trapped components. The propagation constant of
the localized component always lies within the gaps of the linear
spectrum. Therefore this component can be described as a gap
soliton with even or odd symmetry \cite{andrey}, centered at a
maximum or minimum of the grating potential, respectively. Figure
\ref{figsol} shows some of examples of such a gap solitons for
both defocusing and focusing cases. First, we note that the powers
of discrete solitons with different symmetries coincide, i.e.
these solitons belong to the same family. This indicates the
absence of the Peierls-Nabarro potential barrier and good mobility
of the localized states. Second, due to the nonlinear XPM
interaction, the induced grating is strongly modified by the
localized component, but recovers periodicity in the far field.
Different cases of the gap solitons are summarized in
Fig.~\ref{figpow} where we show the families of localized modes
for both focusing and defocusing cases.

In agreement with the theory of gap solitons in nonlinear periodic
structures \cite{andrey}, the families of localized states
originate at the edges of the bands with the numbers $m=2$ (for
$\sigma<0$) or $m=1,3,5...$ (for $\sigma>0$), where the effective
dispersion, $(\partial^2 k_2/\partial K^2)|_{k^m_2}$, is
correspondingly negative or positive. At the respective band edge,
the low-power gap soliton is weakly localized, and can be
described as a slowly varying envelope of the corresponding Bloch
wave $b^m(x)$ \cite{sipe}. In the defocusing case, only gap
solitons [(a) and (b)] can exist in the induced grating, whereas
in the focusing case, both gap [(c) and (d)] solitons and
self-trapped [(e) and (f)] solitons in the semi-infinite gap are
possible.  Near the opposite gap edges, where the gap modes have
high powers, the periodic wave of the grating acquires significant
defects induced by the localized state, however both components
still exist as a vector stationary state.

\begin{figure}
\centerline{\scalebox{0.8}{\includegraphics{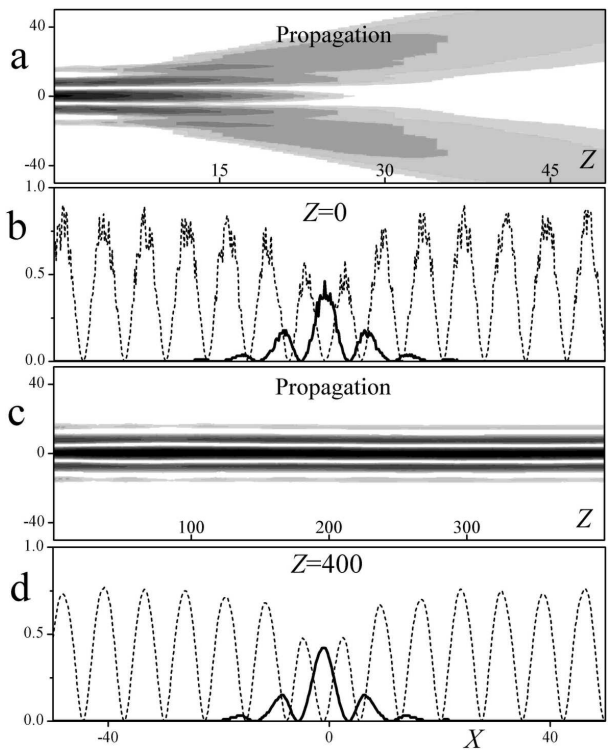}}}\vspace{-1mm}
\caption{Stable propagation of the gap soliton in the defocusing
grating. (a) Propagation dynamics of the initial (odd) state,
corresponding to the family (a-b) in Fig.~\ref{figpow}, perturbed
by a random amplitude noise, in the absence of the grating; (b)
the gap mode (solid) and grating (dashed) initial profiles; (c)
propagation dynamics in the presence of the grating; (d) the final
state at $z=400$. } \label{figmov2}
\end{figure}

To understand the nature of this composite state, we consider the
correction to the linear grating spectrum due to the low-amplitude
gap mode in the $u_2$ component, which is bifurcating off the
lower edge of the band II ($m=3$) in the focusing case, or upper
edge of the band I ($m=2$) in the defocusing case. Near the
bifurcation threshold, the nonlinear XPM coupling leads to the
effective shift of the propagation constant of the periodic
grating component: $k_1=k^g_1+\Delta k_1$, where $k^g_1$ is the
fixed propagation constant for the scalar grating $u^g_1(x)$, and
$\Delta k_1 \sim \sigma \int u^2_2u^2_1(1+u^2_1)^{-2} dx$. As a
result, the band edges $k^m_2$ corresponding to the periodic Bloch
modes $b^m(x)=u^g_1(x)$ shift into the gap. This means that the
nonlinear mode $u_1$, coupled to the field $u_2$ and corresponding
to the fixed propagation constant $k_1=k^g_1$ (i.e. fixed power),
now lies {\em within the band}, and has an {\em oscillatory}, but
not strictly periodic nature. Therefore, the localized low-power
gap mode in the $u_2$ component is nonlinearly coupled to an {\em
in-band} dark-like defect mode in the $u_1$ component, which is
localized on the Bloch-wave background. Together, the two
components form a novel {\em band-gap composite soliton}. In the
{\em deeply nonlinear regime}, when the high-intensity gap mode
induces a large defect in the grating wave, both components exist
as a vector soliton in a mutually formed periodic waveguide
$N(I)$, which now depends on both fields intensities,
$I=u^2_1+u^2_2$. As $k_2 \to k_1$, the gap mode's amplitude
approaches that of the grating, and both components of the
band-gap soliton become oscillatory.

The crucial issue of stability of the gap solitons in the
nonlinear induced gratings is therefore linked to the stability of
the composite band-gap states. We have confirmed dynamical
stability of band-gap solitons by numerical integration of the
vector dynamical model (\ref{NLS}). Figure~\ref{figmov2} shows an
example of the stationary propagation of an odd gap soliton
[family (a-b) in Fig.~\ref{figpow}] in an induced nonlinear
grating for the defocusing case. Both components are initially
perturbed by a random noise at $20\%$ of their peak amplitude
[Fig.~\ref{figmov2} (b)]. If the grating is removed, the localized
gap mode can no longer be supported by the defocusing nonlinearity
and strongly diffracts [Fig.~\ref{figmov2} (a)]. Being coupled to
the lattice, the gap mode generates a defect in the grating and
coexists with it as a dynamically stable composite state, which is
clearly robust to perturbations [Figs.~\ref{figmov2} (c,d)].

In conclusion, we have introduced novel composite band-gap
solitons where one of the components creates a periodic nonlinear
lattice which localizes the other component in the form of a gap
soliton. Nonlinear localization of this kind should be generic to
models of nonlinearly interacting multi-component fields, where
one of the components can exist in a dynamically stable
self-modulated periodic state. Here, we considered a specific
example of a spatial nonlinear photonic structure induced by
optical beams in a photorefractive crystal. Another example is the
dynamical Bragg gratings for optical pulses obtained through the
cross-phase modulation in highly birefringent fibers
\cite{pitois}.

Authors appreciate the critical reading of the manuscript by A.
Sukhorukov. AD gratefully acknowledges support from the Humboldt
Foundation.

\end{document}